\documentclass[12pt]{iopart}

\usepackage{iopams}
\usepackage{graphicx}

\begin{document}

\title[Gravito--magnetic trapping of ${^{87}}$Rb]{Gravito--magnetic trapping of ${^{87}}$Rb}

\author{A Bertoldi\footnote{present address: INFN and Dipartimento di Fisica, Universit\`a di Firenze, I--50019 S.to Fiorentino, Italy} and L Ricci}

\address{Dipartimento di Fisica, Universit\`a di Trento, I--38100 Trento--Povo, Italy}

\ead{ricci@science.unitn.it}

\begin{abstract}
Three--dimensional trapping of neutral atoms in a combined gravito--magnetic potential is reported. Clouds of cold rubidium atoms in different hyperfine states of the ground level were trapped with a lifetime of 4.5 s. Confinement exclusively occured as a combination of the static magnetic force and the gravitational force.
\end{abstract}



\maketitle

\section{Introduction}

Magnetic confinement of neutral atoms is presently a mature technique to produce and study ultracold atomic phenomena. During the last decades, many trapping \cite{Migdall85} and guiding \cite{Friedburg51,Schmiedmayer95,Fortag07} configurations have been devised and experimentally implemented. In these experiments, the magnetic force acting on a species is the dominant confining interaction; additional forces, such as the quadratic Stark interaction \cite{Schmiedmayer03}, are sometimes used to modify the trapping/guiding potential.

In a recent paper \cite{Ricci07}, we have discussed the general theory of combined confinement of neutral atoms. In this new approach, several force fields are tailored so that their combined action results in a confining potential, although each single component, in the absence of the other forces, does not. The forces considered are the magnetic force acting on a paramagnetic species, the diamagnetic force, the quadratic Stark interaction and the standard uniform gravitational field. Key points in using combined potentials for trapping neutral particles are the removal of the constraint of working in a minimum of a single potential --- for example, the trap is located in a region of non--vanishing magnetic field, which prevents confined atoms to undergo Majorana spin--flips ---, and flexibility in designing the confinement potentials. Moreover, the combined approach provides a way to achieve both tight confinement and weak trapping potentials.

In this paper, we describe the experimental implementation of a combined trap for neutral atoms relying on the combination of a magnetic potential and gravity. Cold ${^{87}}$Rb clouds generated with a magneto--optical trap (MOT) \cite{Raab87} are supported against gravity using a precisely tailored magnetic field, which also provides radial confinement. An important aspect regards the role of both force fields: gravity, which usually makes up only a vertical perturbation of the overall potential, is essential here for trapping, as the magnetic potential by itself would not be able to trap the atoms.

The first and, thus far, only achievement of gravito--magnetic trapping of neutral atoms was reported in \cite{Leanhardt03}, where weak three--dimensional confinement allowed to reach extremely low temperature in a condensate. As shown in \cite{Sackett06}, gravity even represents a limiting factor in this experiment: the gravitational field sets a lower bound to the total curvature of the potential well, a constraint that can be overcome by moving to micro-- or null--gravity experiments \cite{Vogel06}. The conditions obtained, namely low temperatures and densities, have thus far been exploited to study quantum reflection \cite{Ketterle04} and to realize interferometry between clouds in the condensate phase \cite{Pritchard05}.

The magnetic field source geometry presented in this letter is compact and cylindrically symmetric. The flexibility in trap design reflects the possibility to easily control the field parameters (eg the curvatures and thus the trap frequencies), leaving the trap minimum location unperturbed. The setup exhibits a sound optical access, a parameter that can be further optimized by means of a planar configuration \cite{Ricci02}. All these features improve the trap used in Ref.~\cite{Leanhardt03}, in which, for example, the trapping region is located only a few mm from a current-carryng wire.

Due to its symmetry, the described gravito--magnetic confinement geometry can be seen as a first step towards the realization of extremely weak trapping potentials, in which the limits set by \cite{Sackett06} can be overcome by exploiting additional, static electric fields.

\section{\label{sec:th} Gravito--magnetic trapping}

Assuming adiabatic evolution of the spin in a non uniform magnetic field, the general form for the gravito--magnetic potential in the case of a paramagnetic low--field seeking (LFS) atom is
\begin{equation}
U_{GM} = U_{G0} + m g z + m_F g_F \mu_B \left | \bi{B} \right | \, , \nonumber
\label{eqn:gravitationalPotential}
\end{equation}
where $U_{G0}$ represents an offset value, $m$ the atom mass, $g$ the modulus of the gravity acceleration, $m_F$ the magnetic quantum number, $g_F$ the Land\'e factor for the atomic state, $\mu_B$ the Bohr magneton, and $\bi{B}$ the magnetic field; the uniform gravitational field is assumed to be directed along $\hat{\bf{z}}$, which is chosen to be the symmetry axis of our cylindrical coordinate system and for the applied magnetic field.

Imposing $U$ to be a confining potential centred in a point requires the gradient of $U$ to vanish and, as a necessary condition, the trace of the Hessian of $U$, or, equivalently, its Laplacian, to be non--negative in the neighborhood of that point. Setting the trap centre at the origin, the condition on the gradient along $\hat{\bf{z}}$ becomes:
\begin{equation}
B' = - \gamma \, , \nonumber \\
\end{equation}
where $B'$ is the axial gradient of the axial component of the magnetic field, evaluated at the origin, and the parameter $\gamma$ is defined as $m g / \mu$, with $\mu \equiv m_F g_F \mu_B$. For the $F=2$, $m_F=2$ state of ${^{87}}$Rb, the value of the parameter $\gamma$ is equal to -15.39 G/cm. This value has to be doubled for atoms either in the $F=2$, $m_F=1$ ($g_F=1/2$) state or in the $F=1$, $m_F=-1$ ($g_F=-1/2$) state.

The gravitational contribution, equivalent to a flat and inclined potential, has a vanishing Hessian. Consequently, only the magnetic contribution defines the confining features of the overall combined potential. Considering the Taylor expansion of the magnetic potential near the origin up to the second order, the axial and the radial curvatures of the combined potential are
\begin{eqnarray}
\frac{\partial^2 U_{GM}}{\partial z^2} = m \omega_z^2 & = & \mu B'' \, , \nonumber \\
\frac{\partial^2 U_{GM}}{\partial \rho^2} = m \omega_\rho^2 & = & \mu \left ( \frac{B'^2}{2B_0} - B'' \right ) \, , \nonumber
\end{eqnarray}
where $\omega_z$ is the axial frequency, $\omega_\rho$ the radial one, and $B_0$ and $B''$ respectively correspond to the axial bias field $B_z$ and the axial field curvature $\partial^2 B_z / \partial z^2$, each evaluated at the origin. The requirement of both curvatures being positive (if a curvature vanishes, higher--order terms have to be taken into account) results in the inequality pair
\begin{eqnarray}
\frac{{B'}^2}{2 B_0} > B'' > 0 \, . \nonumber
\label{eqn:curvatures}
\end{eqnarray}

\begin{figure}
\centering
\includegraphics[width=8.5cm]{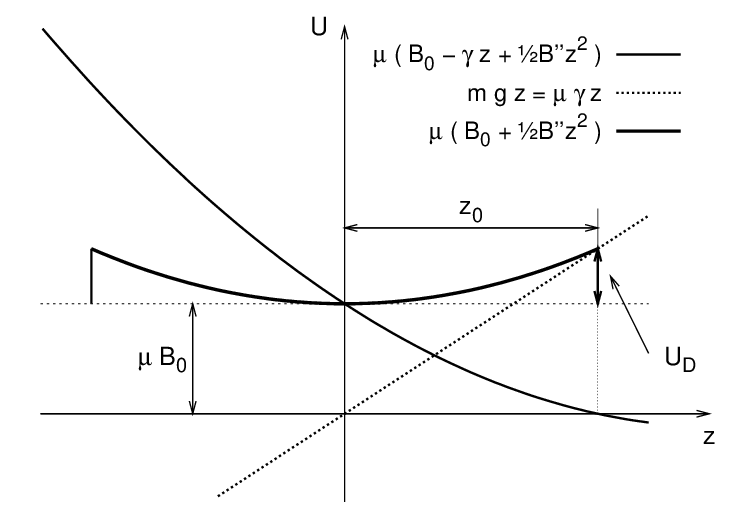}
\caption{\label{fig:depthAndSize} Axial magnetic potential (thin, continuous line), gravitational potential (dashed line; $U_{G0}=0$), and axial combined potential (bold, continuous line) of the gravito--magnetic trap. Size and depth of the resulting potential well are determined by the position $z_0$ of the closest point to the origin where the magnetic field vanishes.}
\end{figure}

The locus of the points where the magnetic field vanishes, allowing the atoms to escape the potential well by flipping their spin, determines depth and size of the confining potential. As a consequence of the requirement of positive radial curvature in the origin, the magnetic field always has at least one zero on the axis. The position $z_0$ of the zero point closest to the origin is:
\begin{equation}
z_0 = \frac{\left | B' \right | - \sqrt{B'^2 - 2 B_0 B''}}{B''} \, . \nonumber \\
\label{eqn:curv}
\end{equation}
As shown graphically in Fig.~\ref{fig:depthAndSize}, the parameter $z_0$ is related to the axial size of the trapping potential and allows for the determination of its depth $U_D$, which is equal to $\frac{1}{2} \mu_B B'' z_0^2$.

\section{Experimental apparatus}
\label{Setup}

The experimental set--up consisted of the vacuum chamber, the laser sources and the related optical system to manipulate the rubidium atoms, and a set of magnetic field sources.

The experiment was realized in a glass cell (external volume $53 \times 55 \times 15$ mm$^3$, glass thickness 2.5 mm) glued on a CF 40 steel flange. The pressure within the cell was of order ${5 \times 10^{-9}}$ mbar.

\begin{figure}[t]
\centering
\includegraphics[width=8cm]{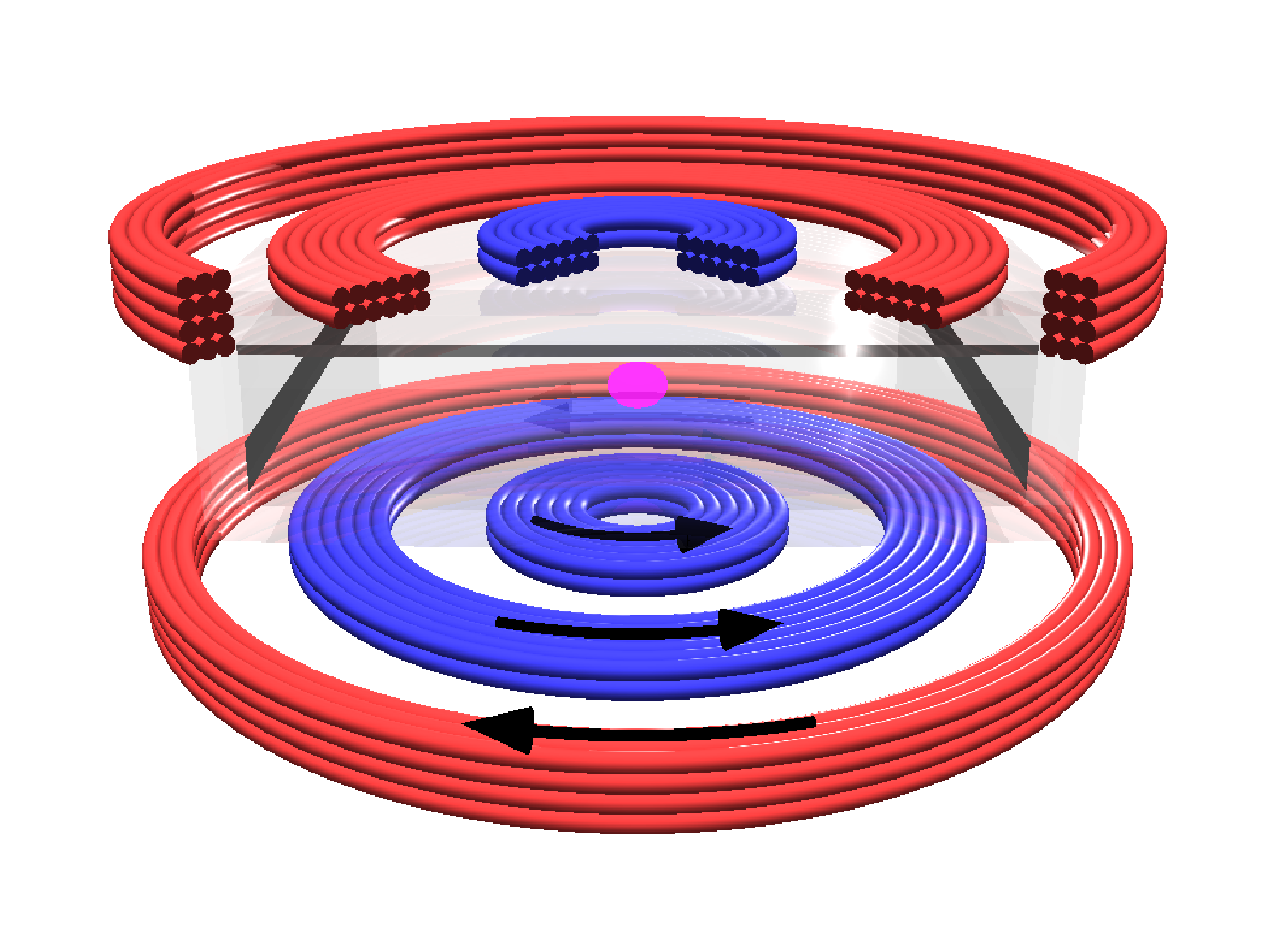}
\caption{\label{fig:setup} Scheme of the experimental apparatus, with the glass cell and the magnetic field sources for the combined trap. The opposite flowing directions for the current are represented by the blue and red colours and, for the bottom coils only, the arrows. The trapping region is shown as a magenta sphere at the configuration centre.}
\end{figure}

A MOT was used as a source of atoms to confine in the gravito--magnetic potential. To optimize the mode--matching, the centre of the MOT and the minimum of the combined trap were imposed to be coincident. For the MOT operation, we used the magnetic quadrupole field characterized by the axial gradient of -15.39 G/cm that compensates the gravitational force in the combined trap for atoms in the $F=2$, $m_F=2$ state. The cooling radiation was generated by a grating--stabilized diode laser \cite{Ricci95}, frequency--stabilized on the ${5 ^2S_{1/2},F=2}$ $\rightarrow$ ${5 ^2P_{3/2},F=3}$ transition of ${^{87}}$Rb by means of a polarization spectrometer \cite{Couillaud80}. Repumping light was provided by another extended cavity diode laser, frequency--stabilized on the $F=1$ $\rightarrow$ $F'=2$ transition of the $D_2$ line. The typical cooling power used for the MOT was 6 mW, distributed over the three counterpropagating beams, each having a diameter of 3 mm. The repumping radiation, with a power of 0.5 mW, was superimposed in one direction only. Under standard operating conditions, $10^6$ atoms were loaded in the MOT with a typical loading time of 3.4 s. The HWHM of the cloud was about 0.5 mm. The atom cloud temperature was further lowered by means of a sub--Doppler cooling phase. The final temperature, measured by turning off the MOT and assessing the expansion of the freely--falling cloud versus time, was typically 6(1) $\mu$K.

\begin{table}
\caption{\label{TaylorExpansion}Main terms of the Taylor expansion of the axial magnetic field in the origin calculated for the three coil pairs, in case of a current of 4.865 A. The sum of the different contributions is also reported.}
\begin{indented}
\item[]\begin{tabular}{@{}cccc}
\br
coil pair & $B_0$ (G) & $B'$ (G/cm) & $B''$ (G/cm)$^2$ \\
\mr
internal & 19.92 & 0 & 91.32 \\
central & 0 & -15.39 & 0 \\
external & -19.48 & 0 & -2.46 \\
\mr
sum & 0.44 & -15.39 & 88.86\\
\br
\end{tabular}
\end{indented}
\end{table}

The adopted magnet configuration was made up of three pairs of coaxial, circular coils as shown in Fig.~\ref{fig:setup}. The coils were placed on two different planes, parallel to the horizontal sides of the glass cell and with a vertical separation of 16.0 mm. The position and size of the six coils were optimized to comply with geometrical and power consumption constraints. The coils were realized using a capton insulated copper wire having a diameter of 1.2 mm. Each internal and central coil was made of $2 \times 6$ windings, with internal diameters of 3.5 mm and 17.0 mm, respectively; each external coil was a $3 \times 4$ windings set with an internal diameter of 30.2 mm. Each pair was used to control one of the three first terms of the Taylor expansion of the magnetic field in the configuration centre: more precisely, the internal, central and external coils were used to control $B''$, $B'$ and $B_0$, respectively. The values of these terms, calculated by integrating the Biot--Savart expression, are reported in Table~\ref{TaylorExpansion}.

The parameter triplet $\{B_0, B', B''\}$, characterizing the overall magnetic field at the origin, was experimentally determined as follows: first, prior to the final assembling, the axial field generated by each coil was measured using a magnetoresistive magnetometer~\cite{Bertoldi05} with an uncertainty of $1 \%$; second, the magnetic field generated by the whole configuration was finely adjusted by using a method that exploits the MOT cloud as a zero magnetic field sensor and allows for the determination of the ratios $B_0/B'$ and $B''/B'$~\cite{bertoldiDissertation}. A residual off--axis component of the bias magnetic field in the combined trap was compensated using two additional coaxial coils oriented along a horizontal axis. This procedure led to the triplet $\{$0.44(1) G, -15.4(4) G/cm, 89(2) G/cm$^2\}$ when the coil setup was fed with a current of 4.865 A. The experimentally determined values agree with calculated values to within $2.5 \%$. Geometric imperfections of the coils, such as tiny displacements of the current-carrying wires from their desired positions, are the main cause for the discrepancies.

The chosen triplet $\{B_0, B', B''\}$ satisfies the confining conditions stated in the previous section for the state $F=2$, $m_F=2$ of ${^{87}}$Rb. At its bottom, the resulting potential well has calculated curvatures corresponding to axial and radial frequencies of $2 \pi \times 12$ Hz and $2 \pi \times 17$ Hz, respectively. Considering the position on the axis where the magnetic field vanishes, $z_0=0.31$ mm, the spin--flip--free trapping region has a depth of about $3~k_B \times \mu$K. The axial and radial size are both about 0.6 mm.

Neglecting the effects of higher order terms in the Taylor expansion of the axial magnetic field --- the third--order contribution $B'''$ is equal to -26 G/cm$^3$ and therefore does not significantly affect the present discussion --- it is easy to show that the axial field reaches a gradient $2 B'$ at $z=-1.7$ mm. The magnetic gradient force in that position exactly compensates gravity for the two remaining trapped states, i.e. $F=2$, $m_F=1$ and $F=1$, $m_F=-1$. In this case, the trap is considerably larger (the axial and radial size are about 3.5 mm and 8 mm, respectively) and deeper (depth of order $60~k_B \times \mu$K). At its bottom, the axial curvature still corresponds to a frequency of $2 \pi \times 12$ Hz, whereas for the radial curvature the corresponding frequency turns out to be about $2 \pi \times 5$ Hz.

To switch between MOT (current only in the gradient coils) and the gravito--magnetic trap (same current flowing in the bias, gradient and curvature coils), an electronic circuit relying on two power MOSFETs was used. The switching was completed within 200 $\mu$s. The different devices as well as the experimental sequence, timing and data acquisition were real--time controlled using a computer running RTAI Linux OS.

\section{Atoms in the combined trap}
\label{Trap}

\begin{figure}
\centering
\includegraphics[width=6.5cm]{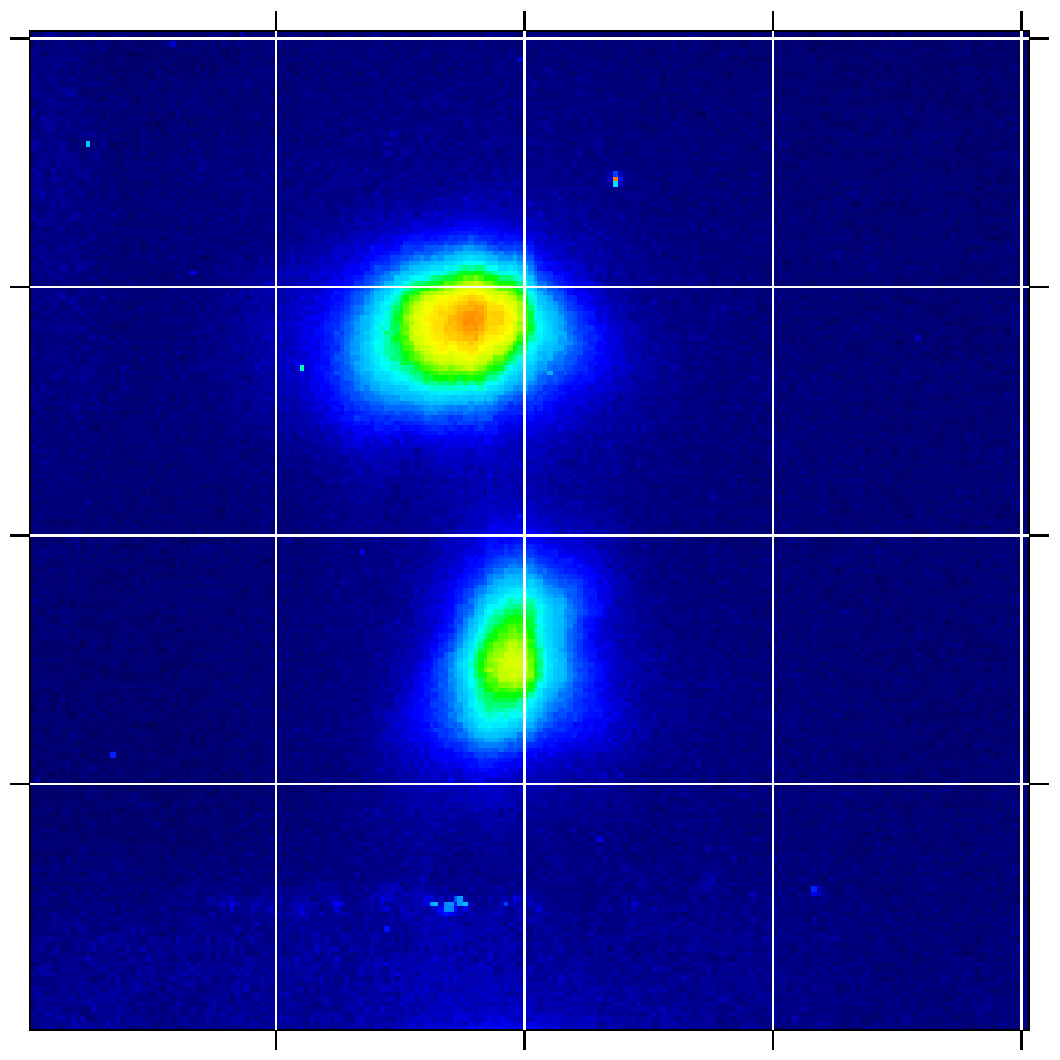}
\caption{\label{fig:double_cloud} Gravito--magnetic trapped clouds of ${^{87}}$Rb atoms in the hyperfine level $F=2$, $m_F=2$ (upper cloud) and $F=2$, $m_F=1$ as well as $F=1$, $m_F=-1$ (lower cloud). The image was taken by shining the MOT cooling beams on the trapping region for 2 ms, after a hold time interval of 150 ms. The distance between the clouds is roughly 2 mm. The upper cloud contains approximately $2 \times 10^4$ atoms.}
\end{figure}

Once prepared in the MOT and cooled down by the optical molasses phase, the atoms were trapped within the combined potential by switching between MOT and the gravito--magnetic trap. At the same time, the laser light shining on the atoms was completely extinguished by the combined action of mechanical shutters and acousto--optic modulators.

After a variable hold time interval, during which free evolution of the atoms in the combined potential occured, the atoms were imaged with resonant light. The MOT beams were shone on the confining region for 2 ms, while the magnetic field was turned off to avoid accelerating the cloud. The fluorescence of the trapped atoms was captured by a high resolution CCD camera (DTA Discovery). Fig.~\ref{fig:double_cloud} shows the fluorescence signal of the atoms in the trapping region after a hold time interval of 150 ms. Atoms transferred from the MOT populated all the hyperfine sub--levels of the ground state. For this reason, two distinct clouds are present: the upper, at the origin, containing atoms in the $F=2$, $m_F=2$ state, and the lower, approximately 2 mm below the former one, containing atoms both in the $F=2$, $m_F=1$ and the $F=1$, $m_F=-1$ state.

To optimize the number of atoms transferred from the MOT to the combined potential, the external magnetic field was compensated within 50 mG using three pairs of independently supplied external coils, which form a cube ($l \simeq 20~$cm) around the vacuum cell. In optimal conditions, about $2 \times 10^4$ atoms in the $F=2$ sub--level were loaded in the gravito--magnetic trap.

Varying the hold time interval up to 10 s, the trapping lifetime was measured to be 4.5(2) s, which is consistent with the background Rb vapor pressure.

Because of the mismatch between the MOT position and the trap bottom, the atoms in the $F=2$, $m_F=1$ and $F=1$, $m_F=-1$ underwent axial oscillations during the hold time interval. Reflecting the nonlinearities intrinsic in the combined potential (primarily due to the dependence of the magnetic contribution on the first power of the modulus of the magnetic field), the dynamics of these oscillations is nonlinear.

\section{Conclusion}

We demonstrated a compact and cylindrically symmetric scheme to realize a combined gravito--magnetic trap for neutral atoms. Cold rubidium atoms in different atomic states were trapped in the combined potential with a lifetime of a few seconds. Further developments of the combined potential technique, possibly relying on the use of the additional contribution given by the Stark interaction, are expected to open a new range of experiments and applications in the experimental framework of trapping and guiding.

\section*{Acknowledgments}

We thank W J Weber for critical reading of the manuscript. This work was partially supported by the INFM Progetto di Ricerca Avanzata ``Photon Matter''.

\end{document}